# Double-excitation manifold's effect on exciton transfer dynamics and the efficiency of coherent light harvesting

Zixuan Hu[1,2], Gregory S. Engel[3], Sabre Kais*[1]

1. Department of Chemistry, Department of Physics, and Birck Nanotechnology Center, Purdue University, West Lafayette, IN 47907, United States
2. Qatar Environment and Energy Research Institute, College of Science and Engineering, HBKU, Qatar
3. Department of Chemistry, James Franck Institute and the Institute for Biophysical Dynamics, University of Chicago, Chicago, IL 60637, United States

*Email: kais@purdue.edu

The efficiency of natural light harvesting systems is largely determined by their ability to transfer excitations from the antenna to the energy trapping center before recombination. Exciton diffusion length similarly limits organic photovoltaics and demands bulk heterojunction architectures. Dark state protection, achieved by coherent coupling between subunits within the antenna, can significantly reduce radiative recombination and enhance the efficiency of energy trapping. In this work we extend the dark state concept to the double-excitation manifold by studying the dynamical flow of excitations. We show the lowest double-excitation state carries minimal oscillator strength but relaxation to this state from higher lying double excitations can be relatively rapid such that the lowest double excitation state can act as a dynamical dark state protecting excitation from radiative recombination. This mechanism is sensitive to topology and operates differently for chain and ring structures, while becoming more pronounced in both geometries when the size of the antenna increases. When the exciton-exciton annihilation (EEA) mechanism is considered, the double-excitation population is quickly depleted and the dynamics changes dramatically. However the efficiency and output power are still significantly different from those calculated using the single-excitation manifold alone, justifying the necessity of considering the double-excitation manifold. Remarkably, in certain scenarios, the EEA can even increase the overall light harvesting efficiency by bringing population down from the double-excitation dark states to the single-excitation manifold.

## I. Introduction

Exciton recombination is an important process to consider in designing high efficiency light harvesting systems[1]. Based on the detailed balance principle, radiative recombination is a fundamental factor to the Shockley-Queisser limit on photovoltaic energy conversion[2]. It has been proposed the detailed balance limit can be broken with coherence induced by either microwave radiation or noise[3,4]. Quantum coherence exists naturally in light harvesting complexes (LHC) that contain multiple chromophores coherently coupled to each other through dipole-dipole interactions[5-17]. In artificial models inspired by natural LHCs, the optically dark states created by quantum coherence effect prevent radiative recombination, effectively increasing the excitation transfer efficiency[18-20]. Theoretical and experimental studies typically examine the single-excitation manifold, and the terms dark and bright states refer to the eigenstates of the Hamiltonian accessible from the ground state via a single excitation[21-23]. A recent study explores the effect of the double-excitation manifold on the energy transfer dynamics for a ring structure with four subunits[24], while expecting diminishing return from increasing the size of the ring. On the other hand, studying the dark state protection mechanism[25], we have observed an abnormal result that increases with the size of the structure, which cannot be explained by the single-excitation manifold alone. In this study we investigate the effect of double-excitation manifold in detail and derive a theory showing the lowest energy double-excitation state can dynamically retain exciton population and potentially enhance energy transfer power by reducing radiative recombination more than expected from the single-excitation dark state alone. This dynamical effect is different for the open chain structure versus the closed ring structure, dependent on the trapping resonance condition, while in both cases the effect grows with the number of subunits in the structure. Exciton-exciton annihilation (EEA) is an important mechanism that quickly transfers excitation from the double-excitation manifold to the single-excitation manifold dissipatively[26, 27]. When EEA is considered, the double-excitation dark state can no longer trap excitation and its effect is reduced. Nonetheless, the efficiency calculated with the EEA included is still significantly different from that calculated with the single-excitation manifold alone, thus reinforcing the importance of the double-excitation manifold. Remarkably, the dissipative process of EEA does not only reduce efficiency as originally expected. In certain scenarios where the excitation can be retained in the double-excitation dark state without a resonance condition necessary for transferring to the trapping site, EEA serves as a channel for the excitation to escape to the single-excitation manifold and then be harvested from there. In such a case the overall efficiency is increased when EEA is introduced.

## II. Model

In this study we use a model described in detail previously[25] wherein an antenna system made of $N$ identical chromophore subunits receives the light energy to generate excitations which are then transferred to a trapping site. The antenna Hamiltonian is:

$$H_a = \omega \sum_{i=1}^{N} \sigma_i^+ \sigma_i^- + J \sum_{i=1}^{N-1} (\sigma_i^+ \sigma_{i+1}^- + \sigma_{i+1}^+ \sigma_i^-) + J_N \left( \sigma_N^+ \sigma_1^- + \sigma_1^+ \sigma_N^- \right) \qquad (1)$$

where $\hbar = 1$, $\omega$ is the site energy, and $J$ is the coupling strength. The use of the Pauli raising and lowering operators explicitly ensures that a single site cannot support more than one excitation. For the ring structure $J_N = J$ while for the chain structure $J_N = 0$. Under Jordan-Wigner transformation the Hamiltonian becomes:

$$H_{JW} = \omega \hat{n} + J \sum_{j=1}^{N-1} \left( c_j^\dagger c_{j+1} + c_{j+1}^\dagger c_j \right) - J_N \left( c_N^\dagger c_1 + c_1^\dagger c_N \right) e^{i\pi \hat{n}} \qquad (2)$$

where $\hat{n} = \sum_j^N c_j^\dagger c_j$ counts the total number of excitations on the ring, $c_j^\dagger$ and $c_j$ are the transformed fermionic creation and annihilation operators. The boundary coupling condition $J_N \left( c_N^\dagger c_1 + c_1^\dagger c_N \right) e^{i\pi\hat{n}}$ in equation (2) changes the sign depending on the parity of the excitation number and the ring structure has a unique double-excitation manifold[28]. For the chain $J_N = 0$ and the solution is independent from the parity of the excitation number. The antenna is connected incoherently to the trapping center through the $N^{th}$ site.

The steady-state dynamics is calculated through a Lindblad optical master equation[24, 25, 29]:

$$\dot{\rho} = -i[H_a + H_t, \rho] + D_o[\rho] + D_p[\rho] + D_{EEA}[\rho] + D_t[\rho] + D_x[\rho] \quad (3)$$

where $\rho$ is the total density operator of the antenna-trap system, $H_t = \omega_t \sigma_t^+ \sigma_t^-$ is the trapping site Hamiltonian. The five Lindblad dissipators -- $D_o[\rho]$, $D_p[\rho]$, $D_{EEA}[\rho]$, $D_t[\rho]$, and $D_x[\rho]$ -- describe four different physical processes.

$D_o[\rho] = \gamma_o \sum_{K',K,\omega_o>0} \Gamma_{K',K} \left( N(\omega_o + 1) \mathcal{D}\left[ \hat{L}_{K',K}, \rho \right] + N(\omega_o) \mathcal{D}\left[ \hat{L}_{K',K}^\dagger, \rho \right] \right)$ is the optical dissipator describing the interband transitions between different excitation levels, where $\gamma_o$ gives the optical transition rate for the antenna, $\Gamma_{K',K} = \left| \langle K' | \sum_{j=1}^N \sigma_j^+ | K \rangle \right|^2$ is the optical coupling strength between two eigenstates of the antenna Hamiltonian with $\omega_o = E_{K'} - E_K > 0$, $N(\omega_o) = \left( e^{\omega_o/k_B T_o} - 1 \right)^{-1}$ is the optical distribution, and $\mathcal{D}\left[ \hat{L}_{K',K}, \rho \right] = \hat{L}_{K',K} \rho \hat{L}_{K',K}^\dagger - \frac{1}{2} \{ \hat{L}_{K',K}^\dagger \hat{L}_{K',K}, \rho \}$ is the Lindblad dissipator with $\hat{L}_{K',K}^\dagger = |K'\rangle\langle K|$.

$D_p[\rho] = \gamma_p \sum_{\mu,\nu,\omega_p>0} \left( N(\omega_p + 1) \mathcal{D}\left[ \hat{L}_{\mu,\nu}, \rho \right] + N(\omega_p) \mathcal{D}\left[ \hat{L}_{\mu,\nu}^\dagger, \rho \right] \right)$ is the phononic dissipator describing the intraband transitions within one excitation level, where $\gamma_p$ gives the phononic relaxation rate, $N(\omega_p) = \left( e^{\omega_p/k_B T_p} - 1 \right)^{-1}$ is the thermal distribution, and $\mathcal{D}\left[ \hat{L}_{\mu,\nu}, \rho \right] = \hat{L}_{\mu,\nu} \rho \hat{L}_{\mu,\nu}^\dagger - \frac{1}{2} \{ \hat{L}_{\mu,\nu}^\dagger \hat{L}_{\mu,\nu}, \rho \}$ where $\hat{L}_{\mu,\nu}^\dagger = |\mu\rangle\langle\nu|$ are the intraband transitions with $\omega_p = E_\mu - E_\nu > 0$.

$D_{EEA}[\rho] = \gamma_{EEA} \sum_{K_2} \mathcal{D}\left[ \hat{L}_{L,K_2}, \rho \right]$ is a phenomenological process describing the EEA process where excitation is transferred from any $|K_2\rangle$ from the double-excitation manifold to the lowest single-excitation eigenstate $|L\rangle$. Similar to the optical and thermal processes,

$$\mathcal{D}\left[\hat{L}_{L,K_2},\rho\right]=\hat{L}_{L,K_2}\rho\hat{L}^\dagger_{L,K_2}-\frac{1}{2}\{\hat{L}^\dagger_{L,K_2}\hat{L}_{L,K_2},\rho\} \text{ and } \hat{L}_{L,K_2}=|L\rangle\langle K_2|.$$ In reality EEA can transfer excitation from any $|K_2\rangle$ to any single-excitation state, but phononic dissipation will quickly bring the excitation to the lowest single-excitation state $|L\rangle$.

$D_t[\rho]=\gamma_t\mathcal{D}\left[\sigma_t^-,\rho\right]$ describes the decay process of the trapping site.

$$D_x[\rho]=\gamma_x\sum_{K_1,\omega_{K_1}=\omega_L}\mathcal{D}\left[\sigma_{K_1}^-\sigma_t^+,\rho\right]\cdot P_N(K_1)+\gamma_x\sum_{K_1,K_2,\omega_{K_2}-\omega_{K_1}=\omega_L}\mathcal{D}\left[\sigma_{K_1}^+\sigma_{K_2}^-\sigma_t^+,\rho\right]\cdot P_N(K_2)$$

describes the incoherent transfer process from the antenna to the trapping site. The first summation is through all the single-excitation states $|K_1\rangle$ with the same energy as the lowest single-excitation state $|L\rangle$ (resonance condition), and $P_N(K_1)$ is the probability of $|K_1\rangle$ projecting into the $N^{\text{th}}$ site. The second summation is through all the double-excitation states $|K_2\rangle$ and all the single-excitation states $|K_1\rangle$ such that their energy difference is equal to $|L\rangle$. In this process an excitation is removed from $|K_2\rangle$ and created in both $|K_1\rangle$ and the trap $|t\rangle$. $P_N(K_2)$ is the probability of $|K_2\rangle$ projecting into the $N^{\text{th}}$ site.

The parameters associated with each of the dissipators in equation (3) are given in Table 1 with the physical meaning of each parameter described in the leftmost panel.

| Parameter | Symbol | Value |
|---|---|---|
| Antenna site energy | $\omega$ | 1.76 eV |
| Antenna coupling strength | $J$ | 100 meV |
| Antenna optical decay rate | $\gamma_o$ | 0.001 meV |
| Antenna phononic decay rate | $\gamma_p$ | 1 meV |
| EEA rate | $\gamma_{EEA}$ | $=\gamma_p$ |
| Antenna-trap transfer rate | $\gamma_x$ | $=\gamma_o$ |
| Trap optical decay rate | $\gamma_t$ | $=\gamma_o$ |
| Ambient temperature | $T$ | 300 K |
| Optical temperature | $T_o$ | 5800 K |

Table 1. Parameters used in the numerical calculations.

In Table 1 the EEA rate is set to be equal to the phononic decay rate. $J=100$ meV is a coupling strong enough such that the thermal distribution and intraband transitions put most excitation population in the lowest eigenstate in each band created by Davydov splitting. In the following we use an H-aggregate (positive $J$) to probe the dark state effect. For the antenna-trap transfer we use $\gamma_x=\gamma_0$ with which the transfer rate is comparable to the interband transitions thus the two processes are competing. This is a different condition from Ref. [25], where a bottleneck condition $\gamma_x=0.01\gamma_0$ was used. In the following we show even with the moderate transfer rate $\gamma_x=\gamma_0$ the dark state mechanism is significant, and the power output is enhanced greatly compared to $\gamma_x=0.01\gamma_0$.

Steady state solution of $\rho$ in equation (3) is obtained using the open-source quantum dynamics software QuTiP[30]. We then use the standard way[4, 24, 31] to calculate the power output of our light harvesting system. The current is determined by $I = e\gamma_t \langle \rho_{te} \rangle_{ss}$, with $\langle \rho_{te} \rangle_{ss}$ being the steady state population of the trap's excited state, $\gamma_t$ being the decay rate of the trap, $e$ being the fundamental charge. The voltage is given by $eV = \hbar\omega_t + k_B T \ln\left(\frac{\langle \rho_{te} \rangle_{ss}}{\langle \rho_{tg} \rangle_{ss}}\right)$, $\langle \rho_{tg} \rangle_{ss}$ being the steady state population of the trap's ground state, $k_B$ the Boltzmann constant, and $T$ the thermal temperature. The total power output for a single antenna-trap system is then simply $P_{out} = IV$.

Here we introduce the concepts of total efficiency and internal efficiency. The total efficiency is defined by $\eta_T = \frac{P_{out}}{P_{Sun}}$ where $P_{out}$ is the total power output for a single antenna-trap system as just described, $P_{Sun}$ is the total light power shined on the system from the sun. $P_{Sun}$ is a complicated quantity dependent on many macroscopic factors such as the material conditions, the sunlight conditions, the atmospheric conditions, etc. In this microscopic study we do not calculate $P_{Sun}$ but instead assume it is constant when the structure of the system changes. This is a reasonable assumption since $P_{Sun}$ is unlikely to change when the microscopic conditions change. When $P_{Sun}$ is constant, $P_{out}$ is then a qualitative indicator of the total efficiency which we report in the unit of picowatt (pW). The internal efficiency is defined by $\eta_I = \frac{P_{out}}{P_{in}}$, where $P_{in}$ is the total power absorbed by the antenna, which is calculated by

$$P_{in} = P_{in0} + P_{in1}$$
$$= \gamma_o \sum_{K_1} \Gamma_{K_1,G} N(\omega_{K_1}) \langle \rho_G \rangle_{ss} \cdot \omega_{K_1} + \gamma_o \sum_{K_2,K_1} \Gamma_{K_2,K_1} N(\omega_{K_2} - \omega_{K_1}) \langle \rho_{K_1} \rangle_{ss} \cdot (\omega_{K_2} - \omega_{K_1}) \quad (4)$$

where $P_{in0}$ is the power input when an optical transition happens between the ground state $|G\rangle$ and the single-excitation state $|K_1\rangle$, $P_{in1}$ is the power input when an optical transition happens between the single-excitation state $|K_1\rangle$ and the double-excitation state $|K_2\rangle$. $\langle \rho_G \rangle_{ss}$ is the steady state population of the ground state $|G\rangle$ and $\langle \rho_{K_1} \rangle_{ss}$ is the steady state population of the single-excitation eigenstate $|K_1\rangle$. The internal efficiency measures by percentage how much energy absorbed by the antenna ends up in the final power output. Since not all the power from the sun can be absorbed by the antenna, the internal efficiency is related to but not the same as the total efficiency. The total efficiency as measured by the total power output is the most important quantity in evaluating the workload of the light harvesting system. On the other hand, the internal efficiency is also an important indicator of how efficient the light harvesting system is. A system with a high total efficiency but a low internal efficiency may power a large workload but generate much wasted power in heat while operating. A system with a low total efficiency but a high internal efficiency may only power a small workload but generate little wasted power in heat as the operation is very efficient.

## III. Results and discussions

We start by numerically calculating both the total power output and the internal efficiency for both the chain and the ring structures.

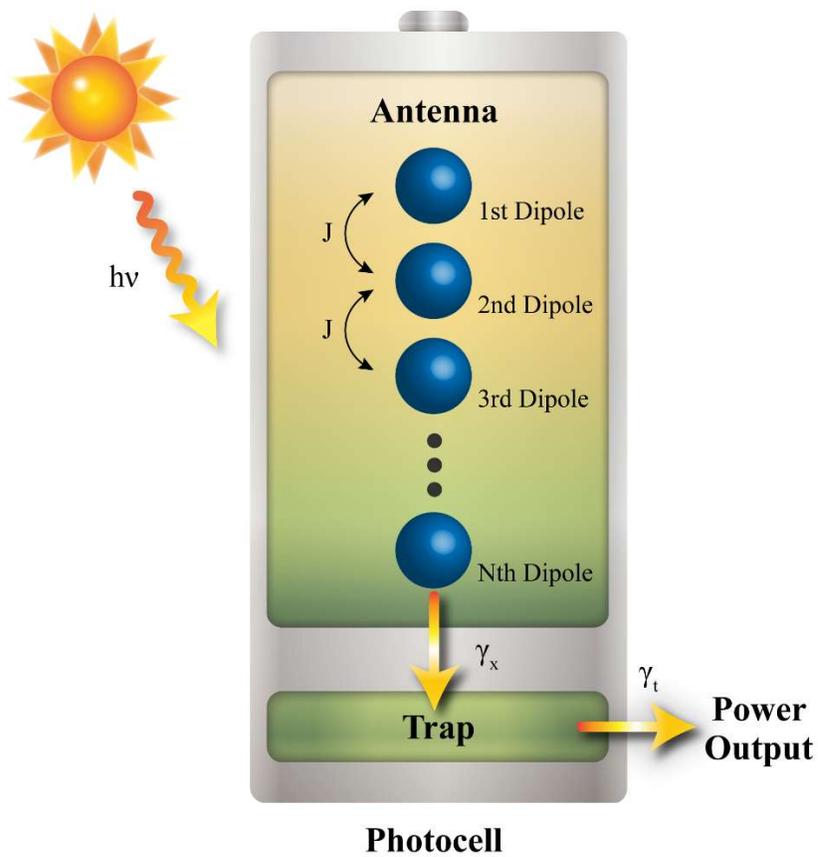

*Figure 1a*

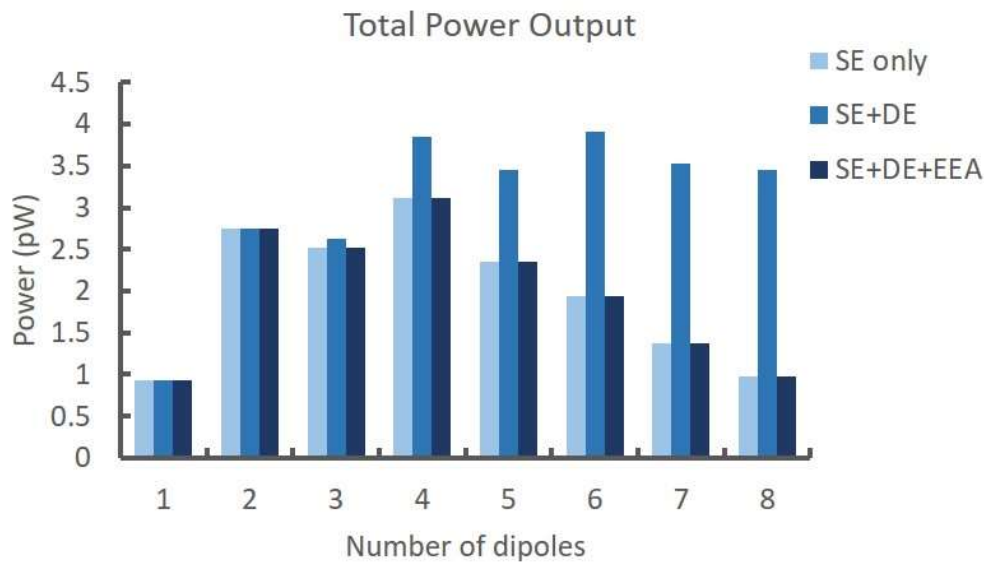

*Figure 1b*

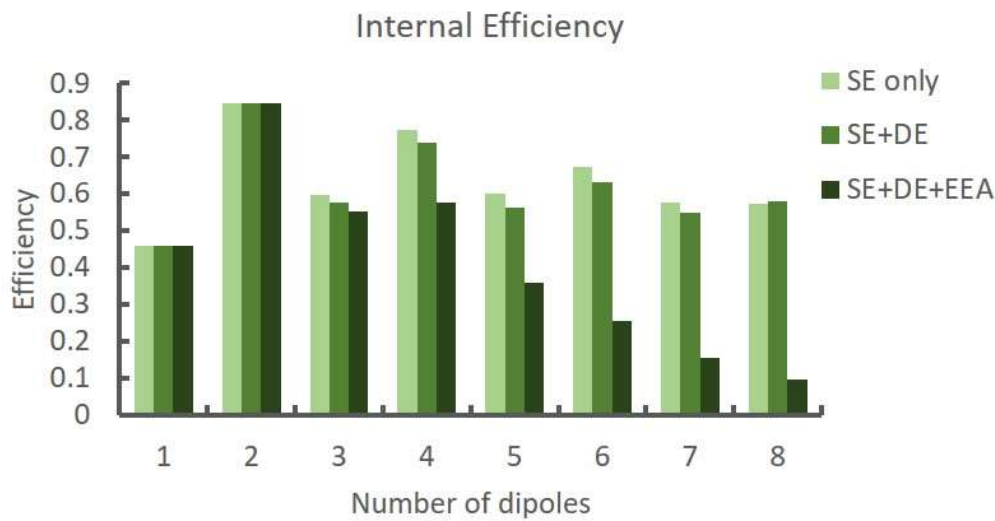

*Figure 1c*

*Figure 1. Showing the results for the chain antenna with the number of chromophores in the antenna ranging from $N = 1$ to $N = 8$. Figure 1a shows the model system with a chain antenna. Figure 1b shows the total power output and Figure 1c shows the internal efficiency. In Figures 1b and 1c for each $N$ we solve the steady state solution of the dynaimcs with 1. the single-excitation manifold alone (SE only, light blue/light green); 2. the single-excitation manifold plus the double-excitation manifold (SE+DE, medium blue/medium green); 3. the single-excitation manifold plus the double-excitation manifold plus the EEA process (SE+DE+EEA, dark blue/dark green). For the total power output, the SE+DE results are significantly greater than both the SE only and*

*SE+DE+EEA results when $N \geq 4$. For the internal efficiency, the SE+DE+EEA results are significantly smaller than both the SE only and SE+DE results when $N \geq 4$*

First for the chain antenna model, in Figure 1b showing the total power output, we can see the power increase from a single chromophore to a two-chromophore chain as well as the zigzag pattern as $N$ increases. These have been presented and explained in Ref. [25] with a bottleneck condition. Their presence here implies the dark state protection mechanism is important even for a moderate antenna-trap transfer rate comparable to the optical transition rate. For $N \geq 4$ the medium blue bar (SE+DE) deviates from the other two and the difference increases as $N$ grows. In Figure 1c showing the internal efficiency, the dark green bar (SE+DE+EEA) starts to deviate from the other two (which have very similar trends) when $N = 4$ and the difference increases as $N$ grows. The differences between the data calculated with different methods show that the double-excitation manifold becomes more important when $N$ is larger. In Figure 1b, the inclusion of the double-excitation manifold greatly enhances the output power while the EEA expectedly removes this enhancement and brings the power output to values similar to those calculated with the single-excitation manifold alone. In Figure 1c, the inclusion of the double-excitation manifold without exciton-exciton annihilation reduces the efficiency slightly, while the EEA reduces the efficiency further because the transition from the single-excitation manifold to the double-excitation manifold adds to the input power in equation (4). With the EEA quickly bringing the excitation back to the single-excitation manifold, there is little power output from the double-excitation manifold. The input power from the single-to-double transition is therefore lost through the EEA (a dissipative process), causing an efficiency loss.

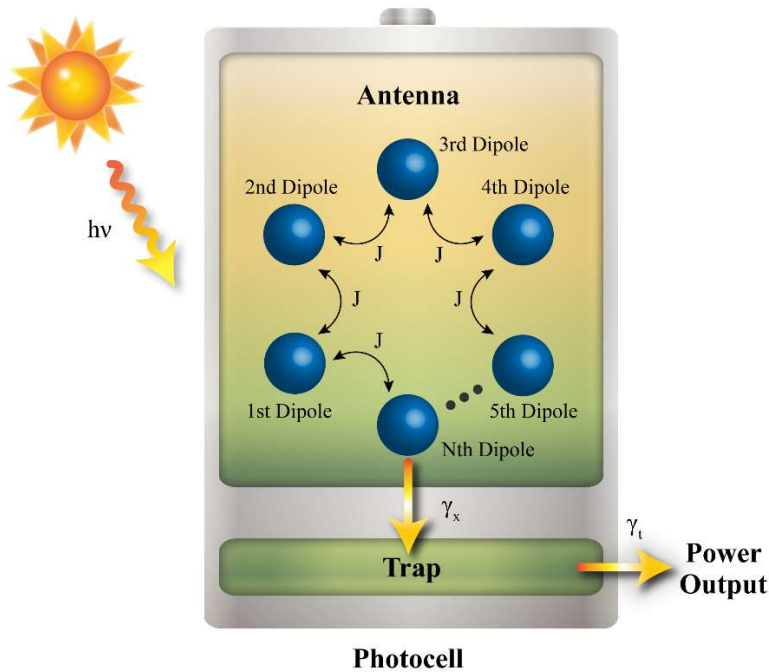

*Figure 2a*

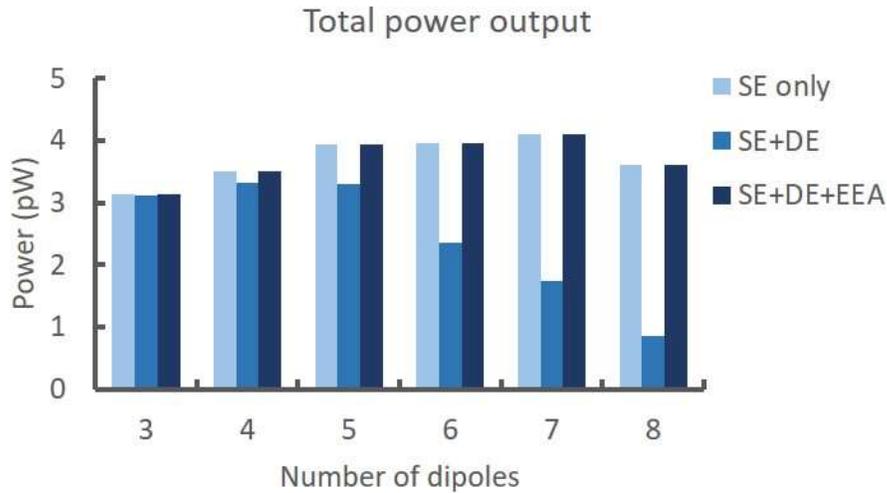

*Figure 2b*

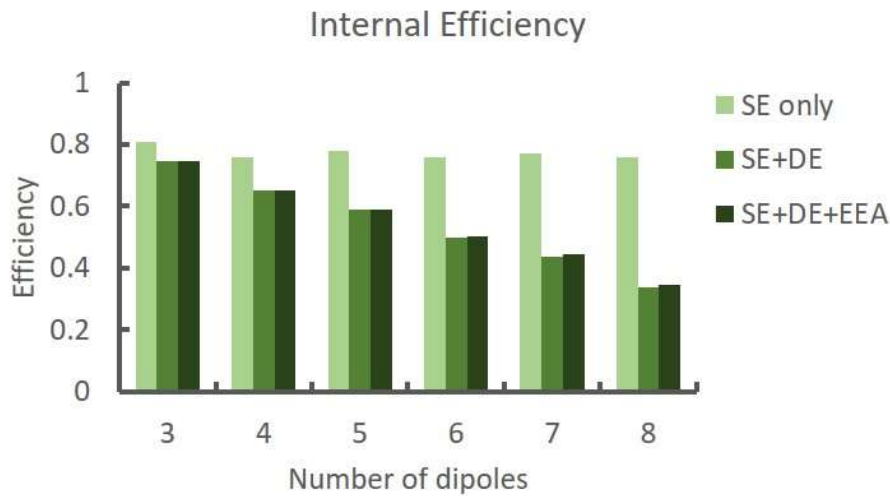

*Figure 2c*

*Figure 2. Showing the results for the ring antenna with the number of chromophores in the antenna ranging from $N = 3$ to $N = 8$ (the smallest ring has to have three subunits). Figure 2a shows the model system with a ring antenna. Figure 2b shows the total power output and Figure 2c shows the internal efficiency. In Figures 2b and 2c for each $N$ we solve the steady state solution of the dynamics with 1. the single-excitation manifold alone (SE only, light blue/light green); 2. the single-excitation manifold plus the double-excitation manifold (SE+DE, medium blue/medium green); 3. the single-excitation manifold plus the double-excitation manifold plus the EEA process (SE+DE+EEA, dark blue/dark green). For the total power output, the SE+DE results are noticeably smaller than both the SE only and SE+DE+EEA results when $N \geq 4$ and this difference increases with $N$. For the internal efficiency, the SE only results are noticeably greater than both the SE+DE+EEA and SE+DE results for all $N$ and this difference increases with $N$.*

Second for the ring antenna model, in Figure 2b showing the total power output, the medium blue bar (SE+DE) starts to deviate from the other two at $N = 4$ and the difference increases as $N$ grows. However the ring antenna is different from the chain antenna in that this time at each $N$ the medium blue bar (SE+DE) is not higher but lower in power than the other two bars. The results in Figure 2c also behave very differently from the chain antenna results, with the two bars with the double-excitation manifold included having almost identical internal efficiency, while the light blue bar (SE only) lying above having almost constant internal efficiency. The EEA's effect is remarkable for the ring, where the inclusion of the double-excitation greatly reduces the power output, while the EEA brings the power up to values similar to those calculated with the single-excitation manifold alone. In other words, the dissipative EEA process does not reduce power output but enhances it. In addition the EEA does not reduce the efficiency compared to results calculated including double-excitation but without EEA. Putting the numerical results together, an important observation is that the inclusion of the double-excitation manifold or the EEA process can produce significantly different power output or internal efficiency, and therefore is necessary for simulating the excitation dynamics. Both the results in Figure 1 and Figure 2 are obtained with the solar temperature $T = 5800 \text{ K}$ under which the steady-state population of double-excitation states is expected to be orders of magnitude smaller than the single-excitation states. Nonetheless we have observed the effects due to the double-excitation manifold for both the chain and the ring structures. In the following we rationalize this surprising fact by analyzing the movement of excitation between the manifolds as shown in Figure 3.

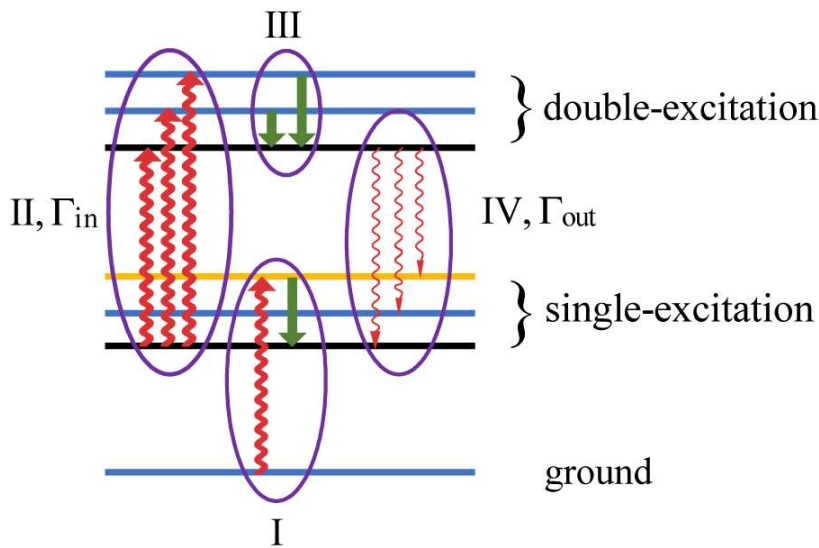

*Figure 3. Four major steps of the dynamical process: I. the initial excitation to the bright state (in yellow) and relaxation to the dark state (in black) on the single-excitation manifold; II. the total incoming flow of excitation from the dark state to the double-excitation manifold; III. the relaxation on the double-excitation manifold to deposit excitation on the lowest double-excitation state (in grey); IV. the total outcoming flow of excitation from the lowest double-excitation state*

*to the single-excitation manifold. In the following we show that Steps II and IV, and particularly the ratio between $\Gamma_{in}$ and $\Gamma_{out}$, determines the tendency with which excitation stay in the double-excitation manifold. When $\Gamma_{in}/\Gamma_{out}$ is large, the lowest double-excitation state serves as a dynamical dark state that can hold a considerable fraction of the excitation population.*

Upon excitation by light, a single excitation is first created at the bright state, from where it will quickly move to the dark state via phononic dissipation. This excitation can be further excited to the double-excitation manifold. The excitation in various double-excitation states will move down by phononic dissipation to the lowest energy state on that manifold, from where it can either move down to the single-excitation manifold or move up to the triple-excitation manifold. In this study we focus on the possibility of moving down only.

The steady-state population of the lowest energy double-excitation state is dependent on three factors: 1. the distribution over intraband levels; 2. the total incoming coupling strength $\Gamma_{in}$ of excitation from the lowest energy single-excitation state to all double-excitation states; 3. the total outcoming coupling strength $\Gamma_{out}$ of excitation to various single-excitation states. The first factor is determined by the Davydov splitting between intraband states. Here we use a coupling strength strong enough ($J = 100$ meV) such that thermal distribution overwhelmingly favors the lowest energy state at room temperature[25]. The second factor is determined by the sum of optical coupling strengths from the lowest single-excitation state $|L\rangle$ to all double-excitation states $|K_2\rangle$. This can be evaluated by deriving a sum rule between the manifolds:

$$\begin{aligned}\Gamma_{in} &= \sum_{K_2} \Gamma_{K_2,L} = \sum_{K_2} \left|\langle K_2 | \sum_i \sigma_i^+ | L \rangle\right|^2 \\ &= \sum_{K_2} \langle L | \sum_i \sigma_i^- | K_2 \rangle \langle K_2 | \sum_i \sigma_i^+ | L \rangle \\ &= \langle L | \sum_i \sigma_i^- \sum_i \sigma_i^+ | L \rangle\end{aligned} \quad (5)$$

where $\Gamma_{K_2,L}$ is the coupling strength from $|L\rangle$ to a double-excitation state $|K_2\rangle$. Considering the coupling from $|L\rangle$ to the ground state $|G\rangle$ is $\Gamma_{L,G} = \left|\langle L | \sum_i \sigma_i^+ | G \rangle\right|^2 = \langle L | \sum_i \sigma_i^+ \sum_i \sigma_i^- | L \rangle$, we obtain:

$$\begin{aligned}\Gamma_{in} - \Gamma_{L,G} &= \langle L | \left[\sum_i \sigma_i^-, \sum_i \sigma_i^+\right] | L \rangle \\ &= \langle L | \sum_i \left(I - 2\sigma_i^+ \sigma_i^-\right) | L \rangle \\ &= N - 2\end{aligned} \quad (6)$$

where we have used the commutation relation $\left[\sigma_\alpha^-, \sigma_\beta^+\right] = \left(I - 2\sigma_\alpha^+ \sigma_\beta^-\right)\delta_{\alpha\beta}$. For the ring structure the lowest single-excitation state $|L\rangle$ is completely dark, $\Gamma_{L,G} = 0$, hence $\Gamma_{in}^{ring} = N - 2$. For the chain structure[25], $\Gamma_{L,G} = \frac{1}{2(N+1)} \cdot \left|\cot\left(\frac{N\pi}{2(N+1)}\right) \cdot \left(1 - (-1)^N\right)\right|^2$, which is zero for even $N$'s and $< 0.1$ for odd $N \geq 3$, therefore $\Gamma_{in}^{chain} \approx N - 2$ with a small error. Hence the total incoming flow of excitation from the lowest single-excitation state to the lowest double-excitation state is approximately $\Gamma_{in} = N - 2$, which increases linearly with $N$. The third factor, the total outcoming flow of excitation to various single-excitation states, is determined by the sum of optical coupling strengths from the lowest double-excitation state to all single-excitation states. For the ring structure we use the selection rule derived in Ref.[28] to find all the allowed transitions from the lowest double-excitation state to the single-excitation manifold and then sum over all the optical coupling strengths. After some algebra, the result is (see the Appendix for detailed analytic derivations):

$$\Gamma_{out} = \begin{cases} \dfrac{2}{N} \tan^2 \dfrac{\pi}{N} & N \text{ odd} \\ \dfrac{4}{N} \tan^2 \dfrac{\pi}{2N} & N \text{ even} \end{cases} \quad (7)$$

where in both cases $\Gamma_{out} \leq N - 2$ (with equality applying only for $N = 3$) and decreases rapidly with increasing $N$. For the chain structure we do not have an analytic form for $\Gamma_{out}$ but numerical results show it has comparable values as in equation (7) and also decreases rapidly with increasing $N$. Now an important quantity is the ratio between $\Gamma_{in}$ and $\Gamma_{out}$, which gives the tendency of the excitation to stay in the double-excitation manifold.

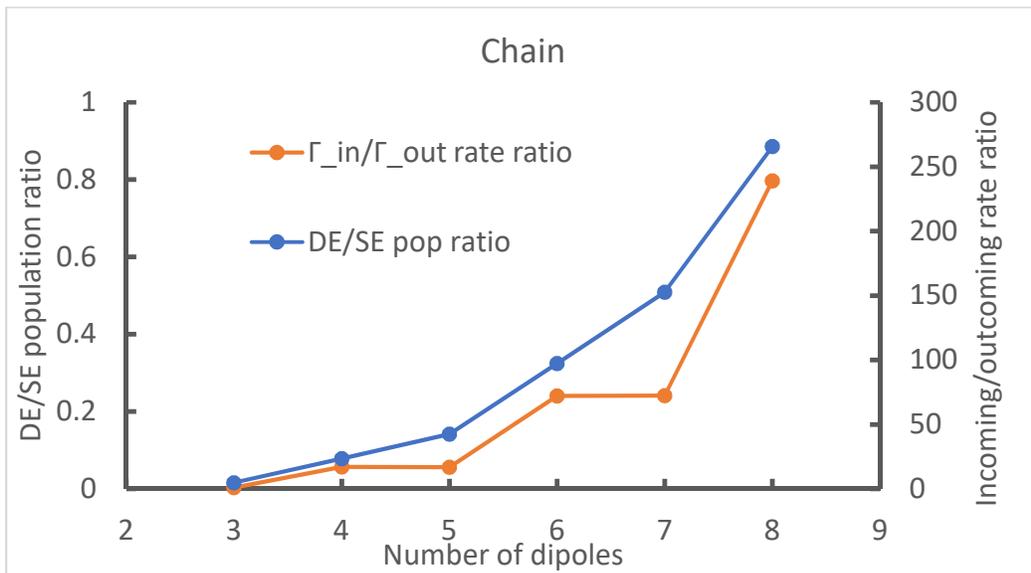

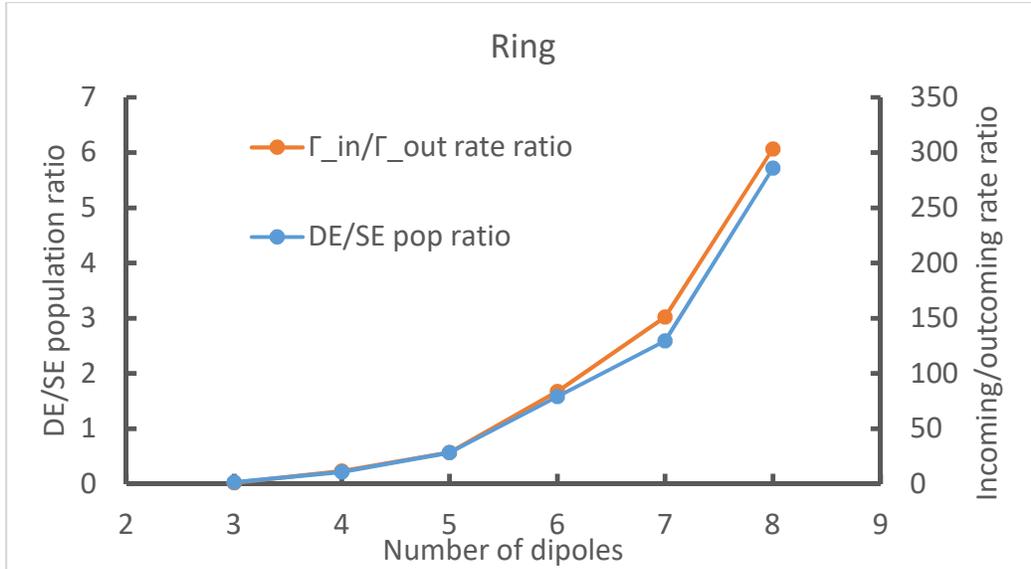

*Figure 4. Showing the incoming/outcoming coupling strength ratio $\Gamma_{in}/\Gamma_{out}$ together with the double-excitation/single-excitation population ratio $DE/SE$. The results are obtained without the EEA process. The two quantities have good correlation for both the chain (upper panel) and ring (lower panel) structures. The results suggest a dynamical dark state concept for the lowest double-excitation state.*

In Figure 4 we plot $\Gamma_{in}/\Gamma_{out}$ against $DE/SE$ -- the ratio between the double-excitation population and the single-excitation population -- without the EEA process. We see a good correlation between the two quantities across different numbers of chromophores. When $N=3$, $\Gamma_{in}/\Gamma_{out}$ is close to unity for both structures. Under the weak solar excitation ($T_o = 5800K$), the double-excitation population is negligible compared to the single-excitation population. Consequently, the double-excitation manifold is not important for $N \leq 3$ in the results shown in Figure 1 and Figure 2. When $N \geq 6$, $\Gamma_{in}/\Gamma_{out}$ is large, and there is considerable population on the double-excitation manifold. Consequently, in Figure 1 and Figure 2 the results for $N \geq 6$ calculated with the double-excitation manifold is significantly different from the ones calculated with the single-excitation alone. The greater double-excitation population at larger $N$ also causes the EEA to have a greater effect.

These results inspire a dark state argument for the double-excitation manifold. When $\Gamma_{in}/\Gamma_{out}$ is large, the lowest double-excitation state is dynamically a dark state in the sense that the total downward transition is not forbidden but much weaker than the total incoming transition from the single-excitation manifold. The double-excitation dark state is therefore an additional important factor in the overall excitation transfer dynamics. For the chain structure in Figure 1, when $N$ is large and the EEA is not considered, the double-excitation dark state can hold excitation population, which enhances the total power output by providing another layer of protection against radiative recombination. The internal efficiency suffers a minor loss because of the intraband transition within the double-excitation manifold. When the EEA is considered, the significant population held on the double-excitation manifold enables the EEA to be an important part of the dynamics. Since the EEA is a very fast process the double-excitation manifold loses the population and the total power output is brought down to the level as calculated with the single-excitation manifold alone. The internal efficiency suffers a major loss because the energy absorbed by the single-to-double transition has not generated any useful power before losing through the EEA, a dissipative process. For the ring structure in Figure 2, when $N$ is large and the EEA is not considered, we still expect the double-excitation dark state to hold excitation population as shown in Figure 4, but this time there is a major loss to the total power output instead of an enhancement. The reason is that the boundary coupling condition $J_N \left( c_N^\dagger c_1 + c_1^\dagger c_N \right) e^{i\pi \hat{n}}$ in equation (2) changes the sign depending on the parity of the excitation number and the ring structure has a unique double-excitation manifold. Specifically the double-excitation states of the ring are not simple composites made from two single-excitation states (like the chain) but are made instead from component states that are calculated with the anti-periodic boundary condition[28]. The consequence of this is the double-excitation states of the ring structure do not have the energy resonance to transfer to the trapping site. In this case the double-excitation dark state functions as a pseudo-trap of excitation without converting it into useful work, and therefore both the total power and the internal efficiency suffer a major loss. When the EEA is considered, the population trapped on the double-excitation manifold can now go back to the single-excitation manifold, and from there gets transferred to the actual trapping site where it is converted to work. In this way the EEA actually improves the total power output as shown in Figure 2, and brings the power up to the level as calculated with the single-excitation manifold alone. The internal efficiency suffers a major loss when the double-excitation manifold is considered, with or without EEA, because either way the energy absorbed from the single-to-double transition is not converted into power output. Combining the results in Figure 1, Figure 2 and Figure 4, we see that 1. the double-excitation manifold is indeed important if we want to have an accurate analysis of the excitation transfer dynamics. 2. The EEA process is important due to the significant population on the double-excitation manifold caused by the dynamical dark state character of the lowest double-excitation state. 3. The total power output and internal efficiency behave very differently when the double-excitation manifold is considered. 4. The chain and ring structures give very different results when the double-excitation manifold is considered, due to the resonance condition for transferring to the trapping site.

## IV.    Conclusion

In this study we have investigated the double-excitation manifold's effect on the exciton transfer dynamics and the efficiency of the coherent light harvesting process. Despite using a weak solar excitation, we show that the inclusion of the double-excitation manifold produces significantly different total power output and internal efficiency compared to those calculated with the single-excitation manifold alone. By analyzing the movement of the excitation between the various manifolds, we propose a concept of dynamical dark state for the lowest double-excitation state, which can hold a considerable fraction of the excitation population on the double-excitation manifold, causing significant changes to the dynamical results. Even in the presence of EEA, the double-excitation manifold can be important to the overall exciton transfer dynamics and must be included for an accurate result for either the total output power or the internal efficiency. Finally, chain and ring structures behave distinctly when the double-excitation manifold and the EEA are considered, due to the resonance condition for the exciton trapping process.

## V. Acknowledgement

Funding for this research was provided by the Qatar National Research Foundation (QNRF), NPRP Exceptional Grant, NPRP X-107-010027.

## VI. Appendix: Analytic derivation for equation (7): the total outgoing optical coupling strength from the lowest double-excitation state to the single-excitation manifold

For the ring structure we solve equation (2) with $J_N = J$. We consider the transitions from the single-excitation manifold to the double-excitation manifold. Single-excitation states have the form:

$$|\psi_k\rangle = \frac{1}{\sqrt{N}} \sum_{j=1}^{N} e^{i\frac{k\pi}{N}j} \sigma_j^+ |0\rangle \tag{8}$$

where $k$ is an even number from $0$ to $2N-2$. Double-excitation states have the form:

$$|\psi_{s_1 s_2}\rangle = C_{s_1}^+ C_{s_2}^+ |0\rangle = \frac{1}{N} \sum_{j<h}^{N} \left( e^{i\frac{\pi}{N}[s_1 j + s_2 h]} - e^{i\frac{\pi}{N}[s_1 h + s_2 j]} \right) \sigma_j^+ \sigma_h^+ |0\rangle \tag{9}$$

where $s_1$ and $s_2$ are odd numbers from $1$ to $2N-1$. The coupling between them is given by:

$$\Gamma_{s_1 s_2, k} = \left| \langle \psi_{s_1 s_2} | \sum_{j=1}^{N} \sigma_j^+ | \psi_k \rangle \right|^2$$

$$= \frac{1}{N^3} \left| \sum_{j<h}^{N} \left( e^{-i\frac{\pi}{N}[s_1 j + s_2 h]} - e^{-i\frac{\pi}{N}[s_1 h + s_2 j]} \right) \left( e^{i\frac{k\pi}{N}j} + e^{i\frac{k\pi}{N}h} \right) \right|^2 \tag{10}$$

$$= \frac{1}{N^3} \left| \sum_{j<h}^{N} \left[ \left( e^{i\frac{\pi}{N}[(k-s_1)j - s_2 h]} - e^{i\frac{\pi}{N}[(k-s_1)h - s_2 j]} \right) + \left( e^{i\frac{\pi}{N}[-s_1 j + (k-s_2)h]} - e^{i\frac{\pi}{N}[-s_1 h + (k-s_2)j]} \right) \right] \right|^2$$

Equation (10) is evaluated to be[28]:

$$\Gamma_{s_1 s_2, k} = \frac{1}{N}\left(\cot\frac{s_2\pi}{2N} + \cot\frac{(s_2-k)\pi}{2N}\right)^2 \tag{11}$$

By the selection rule derived in Ref.[28], a non-zero coupling strength $\Gamma_{s_1 s_2, k}$ requires:

$$s_1 + s_2 - k = 2mN \tag{12}$$

where $m$ is an integer.

Now for $N$ is even, the lowest double-excitation state has $s_1 = N-1$ and $s_2 = N+1$. Since $s_1 + s_2 = 2N$, the only choice of $k$ satisfying equation (12) is $k=0$. Thus to calculate the total coupling strength from the lowest double-excitation state to the single-excitation manifold, we only need to consider one coupling term $\Gamma_{s_1 s_2, k}$ with $s_1 = N-1$, $s_2 = N+1$, and $k=0$, which when substituted into equation (11) gives:

$$\begin{aligned}\Gamma_{out} &= \frac{1}{N}\left(\cot\frac{(N+1)\pi}{2N} + \cot\frac{(N+1-0)\pi}{2N}\right)^2 \\ &= \frac{4}{N}\tan^2\left(\frac{\pi}{2N}\right)\end{aligned} \tag{13}$$

which is the second line in equation (7).

For $N$ is odd, the lowest double-excitation states have $s_1 = N \pm 2$ and $s_2 = N$. By the selection rule in equation (12) there are two finite coupling terms $\Gamma_{s_1 s_2, k}$ that we need to sum over: one with $s_1 = N+2$, $s_2 = N$, $k = 2N-2$ such that $s_1 + s_2 - k = 4N$; the other one with $s_1 = N-2$, $s_2 = N$, $k = 2$ such that $s_1 + s_2 - k = 2N$. Substituting these into equation (11) gives:

$$\begin{aligned}\Gamma_{out} &= \frac{1}{N}\left[\left(\cot\frac{N\pi}{2N} + \cot\frac{(N-2N+2)\pi}{2N}\right)^2 + \left(\cot\frac{N\pi}{2N} + \cot\frac{(N-2)\pi}{2N}\right)^2\right] \\ &= \frac{2}{N}\tan^2\frac{\pi}{N}\end{aligned} \tag{14}$$

which is the first line in equation (7). This concludes the analytic derivation for equation (7).